# Magnetoelectric Control of Domain Walls in a Ferrite Garnet Film

A.S. Logginov, G.A. Meshkov, V.A. Nikolaev, A.P. Pyatakov[*]
Physics Department, M.V. Lomonosov MSU, Moscow, Russia, 119992;
and A.K. Zvezdin
A.M. Prokhorov General Physics Institute, 38, Vavilova st, Moscow, 119991
* Corresponding author: pyatakov@phys.msu.ru

The effect of magnetic domain boundaries displacement induced by electric field is observed in epitaxial ferrite garnet films (on substrates with the (210) crystallographic orientation). The effect is odd with respect to the electric field (the direction of wall displacement changes with the polarity of the voltage) and even with respect to the magnetization in domains. The inhomogeneous magnetoelectric interaction as a possible mechanism of the effect is proposed.
**DOI:** 10.1134/S0021364007140093

The last few years marked the great progress in the field of magnetoelectric materials [1-3]. The newly discovered ferroelectricity induced by spiral magnetic ordering [3-7] and early discovered inverse effects of electrically induced spin modulation [8-10] not only provide a deeper insight into the mechanisms of magnetoelectric coupling (the so-called inhomogeneous magnetoelectric interaction [9-11]) but also map out the route for electric control of micromagnetic structure in solids, the possibility predicted in 1980-ies [11] and still not realized.

The best candidates for the specific character of micromagnetism in magnetoelectric materials are thin films of ferrite garnets. From the one hand they are classical object to study micromagnetism, [12-15] from the other hand they exhibit a magnetoelectric effect that is an order of magnitude greater than the corresponding effect in the classical $Cr_2O_3$ magnetoelectric [16]. Electromagneto-optical effects on local areas of a ferrite-garnet film revealed that effect is vanishingly small in a homogeneously magnetized film but it increases drastically in the vicinity of domain walls. This effect was attributed to "breathing" of domain wall in electric field providing indirect evidence for the influence of electric field on micromagnetic structure [17].

In this Letter we report on the direct magnetooptical observation of a new manifestation of the magnetoelectric effect, namely, the electric-field-control displacement of domain walls in ferrite garnet films, that is of interest from the fundamental point of view and opens up new possibilities for the development of multipurpose spintronic and magnetophotonic devices on a single material platform.

In our experiments we used the 9.7-μm-thick epitaxial $(BiLu)_3(FeGa)_5O_{12}$ ferrite garnet films grown on a $Gd_3Ga_5O_{12}$ substrate with the (210) crystallographic orientation. The substrate thickness was about 0.5 mm. The period of the strip domain structure was 34.5μm, and the saturation magnetization was $4\pi M_s$= 53.5G. To produce a high-strength electric field in the dielectric ferrite garnet film, we used a 50μm-diameter copper wire with a pointed tip, which touched the surface of the sample (Fig. 1). The diameter of the tip of the copper "needle" was about 20 μm. This allowed us to obtain an electric field strength of up to 1500 kV/cm near the tip by supplying a voltage of up to 1500 V to the needle. The field caused no dielectric breakdown, because it decreased rapidly with distance from the needle and, near the grounding electrode (a metal foil attached to the substrate), did not exceed 600 V/cm. The absence of the possible leakage currents between the needle and the grounding electrode (e.g., over the sample surface) was verified by a milliampermeter. To observe the domain structure, the polarization method based on the Faraday effect was used. For the observation in transmitted light, a hole ~ 0.3 mm in diameter was made in the grounding electrode. The image of the magnetic structure was obtained using a CCD camera connected with a personal computer.

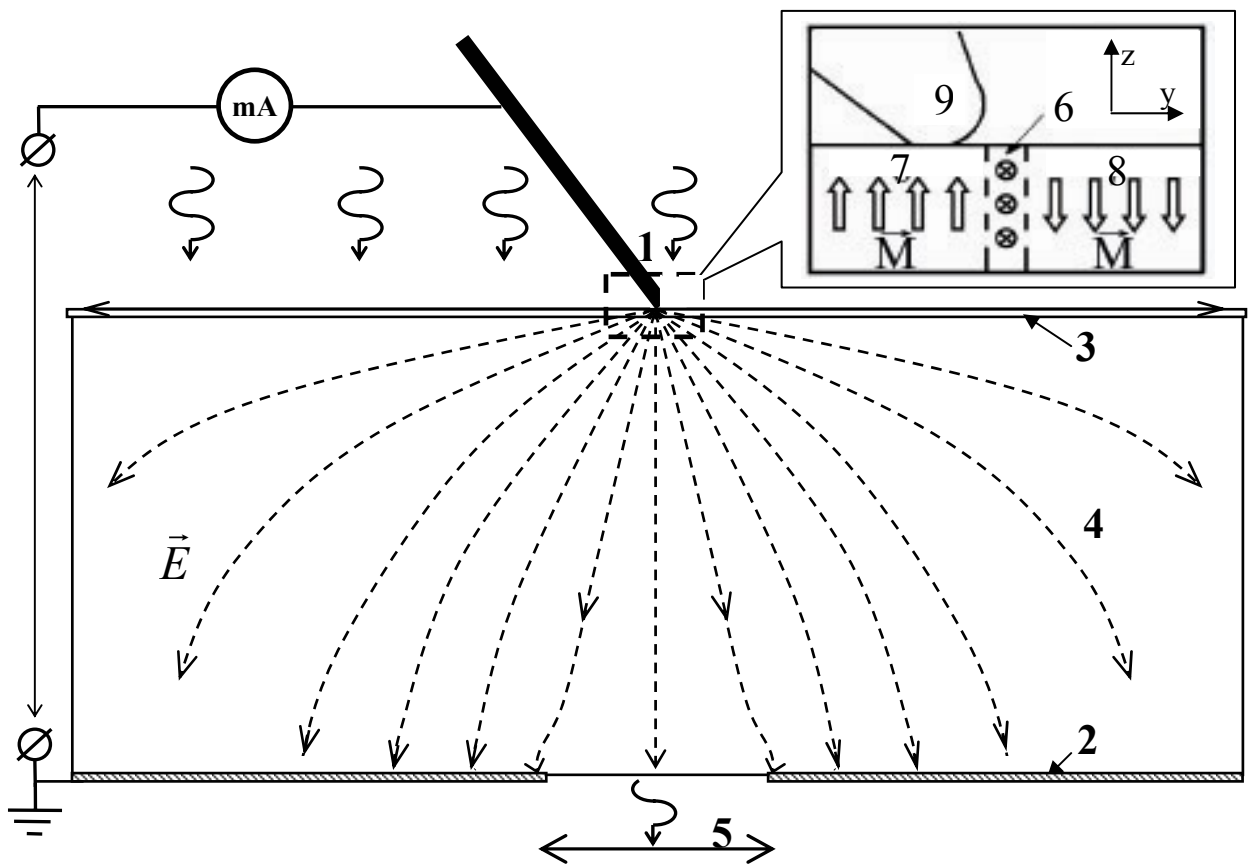

**Fig. 1** Schematic representation of the geometry of the experiment and the configurations of the electric field and magnetization. The electric field (the field lines are shown by the dashed lines) is formed in the dielectric medium of the sample between the needle (*1*) and the metal foil (*2*),which plays the role of the grounding electrode. The maximum field strength (above 1 MV/cm) is reached in the magnetic film (*3*) near the tip; it decreases rapidly in the bulk of the substrate (*4*) and does not exceed 600 V/cm near the grounding electrode (*2*). The abscence of the leakage currents is controled with the milliampermeter (mA). The light (beams are denoted with wavy arrows) is incident along the normal to the surface. The objective lens (*5*) is placed behind the pinhole in the foil (*2*). The inset shows the magnetization distribution in the film: the domain wall (*6*) separates two domains (*7, 8*) with opposite magnetization directions; the tip (*9*) touches the ferrite garnet surface near the domain wall.

In the experiment, we measured the static distribution of magnetization before and after the electric field was switched on. As a result, we obtained pairs of images for different voltage polarities and different needle positions. In all of the series, when a dc voltage was applied between the needle and the substrate, we observed a local displacement of the domain wall near the tip (Fig. 2). The magnitude of the displacement increased with voltage.

We found three characteristic features of the phenomenon, which serve as the basis for our subsequent consideration.

(i) The direction of the domain wall displacement depends on the polarity of the voltage (and, hence, on the direction of the electric field): in the case of positive polarity, the wall was attracted to the needle, and, in the case of negative polarity, it was repulsed.

(ii) The direction of the wall displacement did not depend on the direction of magnetization in the domain (along the $z$ axis or against it, see the inset in Fig. 1).

(iii) The magnitude of the effect increased strongly with a decrease in the distance between the tip and the domain wall. The most pronounced effect was observed at a positive voltage, when the domain wall was pulled into the region of high electric field strength in the immediate vicinity of the point of contact of the needle and the surface (Fig. 2).

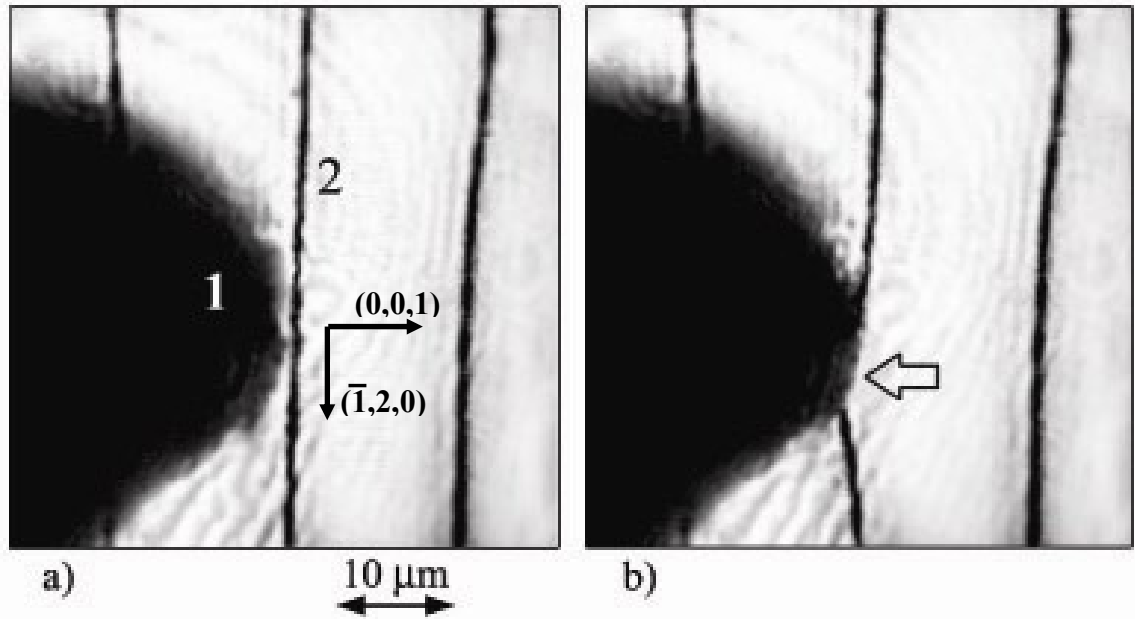

**Fig. 2** The effect of electric field in the vicinity of electrode (1) on magnetic domain wall (2) in the films of ferrite garnets: a) initial state b) at the voltage of +1500 V applied

The characteristic features listed above allow us to exclude the effects of non-magnetoelectric nature that could lead to displacements of domain walls: the magnetic fields caused by possible leakage currents and the magnetostrictive phenomena caused by the pressure of the tip on the sample due to electrostatic attraction. Indeed, the dependence on the polarity of the voltage applied to the needle (feature (i)) allows us to exclude the effect of the tip pressure on the sample, because the tip polarizes the sample surface and is attracted to it irrespective of the sign of the potential at the needle; hence, the effect caused by the tip pressure should be independent of polarity. Feature (ii) testifies that, even in the presence of leakage currents, the effect cannot be related to the magnetic field generated by the electric currents, because, otherwise, the domain wall would be displaced in opposite directions for domains with opposite magnetization directions. Thus, features (i) and (ii) of the phenomenon under study allow us to conclude that the latter is of magnetoelectric nature.

Concerning feature (i), it should be emphasized that because of the presence of an inversion center in the crystal symmetry group of *bulk* ferrite garnet samples only the effects that are proportional to even powers of electric field are possible, which manifests itself in quadratic magnetoelectric [18] and electromagnetooptic [19] effects. The dependence of the direction of the domain wall displacement on the electric polarity (the oddness of the effect with respect to electric field) testifies to the violation of the spatial inversion in *films*, unlike the case of bulk ferrite garnet samples.

The third feature (iii), namely, the sharp increase in the magnitude of the effect observed when the domain wall is pulled closer to the tip, indicates that the electric field selectively acts on the regions with inhomogeneous magnetization, i.e., the domain walls. Indeed, in this case, the domain wall approaches the point of the needle-surface contact and comes into the region of high electric field strength (Fig. 1). If we dealt with a homogeneous magnetoelectric effect, the electric field would also act upon the regions of homogeneous magnetization, i.e., domains, by decreasing (or increasing) the domain over which the tip is positioned. In this case, one should expect an opposite behavior of the magnitude of the effect as a function of the tip–domain wall distance, because a displacement of the wall toward the needle would bring the neighboring domain into the region of high electric field strength as well. The fact that the needle selectively acts on the domain walls suggests that the phenomenon observed is a manifestation of the inhomogeneous magnetoelectric effect related to micromagnetic inhomogeneities of the material [11,20].

$$F_L = \gamma_{ijkl} \cdot E_i \cdot M_j \cdot \nabla_k M_l, \qquad (1)$$

where **M=M(r)** is magnetization distribution, **E** is electric field, $\nabla$ is vector differential operator, $\gamma_{ijkl}$ is the tensor of inhomogeneous magnetoelectric that is determined by the symmetry of the crystal. One can learn imediately from the equation (1) that the effect is odd in electric field **E**, and doesn't change the sign with magnetization **M** reversal, that agrees with the main features of the effect. It follows from (1) that effect is selective to the magnetic inhomogeneities, i.e. it influences on the domain walls rather than domain itselves. This property agrees with the enhancement of the effect observed when the wall was drawn in the region of the contact where the electric field strength was maximal.

The influence of the inhomogeneous magnetoelectric effect on the domain walls is the stronger, the greater components of magnetoelectric interaction tensor $\hat{\gamma}$ and the smaller the characteristic size of magnetic inhomogeneity $\Delta$ are in the given material. The components of tensor $\hat{\gamma}$ can be estimated from Eq. (1) and from the condition of the equality of the magnetoelectric and magnetostatic energies with the use of the known magnetic field parameters and experimental data: the saturation magnetization $Ms$= 53.5 G, the domain wall width $\Delta$= 100 nm, the electric field $E$=1MV/cm = 3.3 $10^3$CGS, and the volume density of magnetostatic energy $F_{m-st}$=0.1 erg/cm$^3$ (estimated from the deflection of the domain wall) [12]. The value obtained for the inhomogeneous magnetoelectric interaction tensor components are $\gamma \sim 10^{-9}$ CGS (for comparison, a similar constant for the effect that gives rise to spatially modulated spin structures in bismuth ferrite [10] is $\gamma=10^{-11}$ CGS).

Despite the fact that all the features of the effect testify that the observed phenomenon is caused by the inhomogeneous magnetoelectric interaction in the ferrite garnet material [11,20], we cannot completely exclude the mechanism determined by the local anisotropy variation due to mechanical stress associated with the piezoelectric effect. To distinguish between these mechanisms with better accuracy, additional studies are necessary, for example, the study of the effect as a function of the characteristic size of micromagnetic inhomogeneities and dependence on the direction of the electric field.

We are grateful to A.V. Khval'kovskii for the interest in our study and for valuable discussions. This work was supported in part by the Russian Foundation for Basic Research (project no. 05-02-16997) and the "Dynasty" Foundation.

1. Manfred Fiebig, "Revival of the magnetoelectric effect", *J. Phys. D: Appl. Phys*. **38**, R123–R152 (2005)
2  W. Eerenstein, N. D. Mathur & J. F. Scott,  Multiferroic and magnetoelectric materials, Nature, **442,** 759 (2006)
3. S.-W. Cheong, M. Mostovoy, Multiferroics: a magnetic twist for Ferroelectricity, Nature Materials, **6**, 13 (2007)
4. Maxim Mostovoy, Ferroelectricity in Spiral Magnets, PRL, **96**, 067601 (2006)
5. A. M. Kadomtseva, Yu.F. Popov, G.P. Vorob'ev, K. I. Kamilov, A. P. Pyatakov, V. Yu. Ivanov, A. A. Mukhin, A. M. Balbashov, Specifity of magnetoelectric effects in new GdMnO$_3$, magnetic ferroelectric, JETP Letters, **81**, iss.1, pp.19-23 (2005)
6. Yoshinori Tokura, Multiferroics as Quantum Electromagnets, Science, **312** , 1481 (2006)
7. E. V. Milov, A. M. Kadomtseva, G. P. Vorob'ev, Yu. F. Popov, V. Yu. Ivanov, A. A. Mukhin and A. M. Balbashov, Switching of spontaneous electric polarization in the DyMnO$_3$ multiferroic, **85**, p. 503 (2007)

8. I. M. Vitebskii, Sov.Phys. JETP, **55**, 390 (1982)
9. Bar'yakhtar, V.G., and Yablonskiy, D.A. Formation of long-period structures in orthorhombic an rhombohedral antiferromagnets in applied fields. Sov. Phys. Solid State, **24**,1435 (1982)
10. A. Sparavigna, A. Strigazzi, A.K. Zvezdin, Electric-field effects on the spin-density wave in magnetic ferroelectrics, Phys. Rev. B, **50**, 2953 (1994)
11. Bar'yakhtar, V.G., L'vov, V.A., and Yablonskiy, D.A. Inhomogeneous magneto-electric effect. JETP Lett.; **37**, 673 (1983)
12. A. P. Malozemoff and J. C. Slonczewski, Magnetic Domain Walls in Bubble Materials, Academic, New York, 1979.
13. A.K.Zvezdin, V.A.Kotov, Modern magnetooptics and magnetooptical materials, IOP Publising, UK, 400 p, 1997
14. M. V. Chetkin, V. B. Smirnov, A. F. Popkov, et al., Sov. Phys. JETP **67**, 2269 (1988)
15. A. S. Logginov, A. V. Nikolaev, V. N. Onishchuk, and P. A. Polyakov, JETP Lett. **66**, 426 (1997).
16. B. B. Krichevtsov, V. V. Pavlov, and R. V. Pisarev, JETP Lett. **49**, 535 (1989).
17. V. E. Koronovskyy, S. M. Ryabchenko, and V. F. Kovalenko, Electromagneto-optical effects on local areas of a ferrite-garnet film, Phys. Rev. B **71**, 172402 (2005)
18. T. H. O'Dell, Philos. Mag. **16**, 487 (1967).
19. B. B. Krichevtsov, R. V. Pisarev, and A. G. Selitskii, Sov. Phys. JETP Lett. **41**, 317 (1985)
20. A.S. Logginov, G.A. Meshkov, A.V. Nikolaev, A. P. Pyatakov, V. A. Shust, A.G. Zhdanov, A.K. Zvezdin, Electric field control of micromagnetic structure, Journal of Magnetism and Magnetic Materials, **310**, 2569 (2007)